\documentstyle[11pt,psfig]{article}
\newcommand{\vph}{\varphi}

\newcommand{\lar}{\leftarrow}

\newcommand{\Rar}{\Rightarrow}
\newcommand{\n}{\mbox{\bf{not}}}

\newcommand{\Dn}{{\mathrm DATALOG}^{\neg}}
\newcommand{\sol}{{\mathit sol}}

\newtheorem{theorem}{Theorem}[section]

\newtheorem{example}{Example}[section]
\newtheorem{definition}{Definition}[section]

\newtheorem{remark}{Remark}[section]
\newtheorem{theorem1}{Theorem}[subsection]

\newtheorem{example1}{Example}[subsection]
\newtheorem{definition1}{Definition}[subsection]

\newtheorem{remark1}[theorem1]{Remark}

\begin{document}
\ \\ 
\ \\
\begin{center}
{\Large\bf Stable models and an alternative logic programming paradigm}

\ \\
\ \\
{\it Victor W. Marek, Miros\l aw Truszczynski}
\ \\
Department of Computer Science\\
University of Kentucky\\
Lexington, KY 40506-0046\\
{\tt \{marek,mirek\}@cs.engr.uky.edu}
\end{center}
\ \\
\begin{abstract}
In this paper we reexamine the place and role of stable model semantics
in logic programming and contrast it with a least Herbrand model
approach to Horn programs. We demonstrate that inherent features of stable 
model semantics naturally lead to a logic programming system that
offers an interesting alternative to more traditional logic programming
styles of Horn logic programming, stratified logic programming and 
logic programming with well-founded semantics.
The proposed approach is based on the interpretation of program clauses as 
constraints. In this setting programs do not describe a single intended 
model, but a {\em family} of {\em stable} models. These stable models 
encode solutions to the constraint satisfaction problem described by 
the program. Our approach imposes restrictions on the syntax of logic 
programs. In particular, function symbols are eliminated from the language. 
We argue that the resulting logic programming system is well-attuned to 
problems in the class NP, has a well-defined domain of applications, 
and an emerging methodology of programming. We point out that what makes 
the whole approach viable is recent progress in implementations of algorithms 
to compute stable models of propositional logic programs.
\end{abstract}

\ \\    
\section{Introduction}
\label{intro}

Stable model semantics appeared on the logic programming scene in the
late 80s in an effort to provide an understanding of programs with
negation. Since it was first proposed by Gelfond and Lifschitz
\cite{gl88}, it was regarded by  logic programming community with a
dose of reserve and unease. It was intuitively felt that the stable 
model semantics properly deals with negation and some formal evidence 
supporting this intuition was established. At the same time, the stable 
model semantics did not fit into a standard paradigm of logic programming 
languages. While standard approaches assign to a logic program a single 
``intended'' model, stable model semantics assigns to a program a
{\em family} (possibly empty) of ``intended'' models. Further, in the
presence of function symbols, Horn logic programs can specify any 
recursively enumerable set, while stable 
model semantics increases the expressive power of logic programs well 
beyond acceptable notions of computability. Finally, the SLD-resolution, 
the true bread-and-butter of logic programmers and the heart of control 
mechanisms behind standard logic programming implementations, seems to be 
inappropriate for the stable model semantics. 

As a consequence of these difficulties in reconciling the stable model
semantics with a traditional paradigm of logic programming, the stable 
model semantics received relatively less attention from the logic 
programming community than other semantics proposed for programs with
negation such as perfect model semantics for stratified programs and
well founded semantics. 

In this paper we argue that rather than to try to resolve these
inconsistencies and force stable model semantics into a standard 
logic programming mold (this effort most likely is doomed to failure), 
a change of view is required. Therefore, we propose a perspective 
on the stable model semantics that departs from several basic tenets of 
logic programming. At the same time, this perspective leads to a
computational system very much in the general spirit of logic
programming. The system is declarative, retains the separation of logic
from control, has a well-defined domain of applications and 
emerging programming methodology. We refer to this version of 
logic programming as {\em stable logic programming} (or, SLP, for 
short).

There are several key elements to the view on the stable model semantics
that we describe here. First, we restrict the syntax by disallowing function
symbols. Thus, the syntax of SLP is the same as the syntax of DATALOG
with negation. The restriction of the syntax has significant 
effect on the expressive power of programs. In particular, it 
curtails the ability to use recursion.

Second, we view a program as specifying a collection of models rather 
than a single model. Thus, SLP seems to be especially well suited for
all these problems where solutions are subsets of some universe, as each
solution can be modeled by a different stable model. Many combinatorial
and constraint satisfaction problems fall into this category.

The restricted syntax (limiting the use of recursion) and the shift 
in the semantics (programs specify a collection of models rather than 
a single model), change the way in which we interpret and design programs. 
Programs are interpreted as sets of constraints. Clauses represent 
individual constraints on objects of interest rather than recursive 
definitions of these objects. Hence, a different approach to programming 
is needed. Objects that in Horn logic programming would be represented 
as terms and defined recursively, in SLP are represented as different 
stable models and defined in terms of constraints. 

Finally, although the SLP follows the basic logic programming tenet 
of ``uniform control'', the control used in SLP is different. Instead 
of SLD resolution used in Horn Logic programming, a backtracking search 
for stable models is used.

While (Horn) logic programming is well attuned to the concepts of
Turing computability, recursively enumerable sets and partial recursive
functions, SLP is related to a much more narrow class of
problems. As we point out, all decision problems that are in NP can be 
solved within the paradigm of SLP. Moreover many (possibly all) search 
problems, whose decision versions are in NP, can be solved with SLP
programs, too. While NP is much smaller than the class of r.e. sets, it 
still covers a wide collection of important computational
problems including many combinatorial optimization and constraint
satisfaction problems. This is one of the main reasons why we believe
that SLP can evolve into a useful computational tool.

Recently, the idea of using logic programs with restricted syntax and
nonstandard semantics was discussed in several papers. Niemel\"a
\cite{nie98}, in a closely related work, proposed function symbol-free logic 
programming with stable model semantics as a vehicle to process constraint 
satisfaction problem. Second, Cadoli and others \cite{cp98a}, proposed
the use of DATALOG programs {\em without} negation and with the semantics of
parallel circumscription as a tool for solving decision problems in NP.
Finally, in \cite{cmmt98} we study a similar use of default logic 
(a formalism extending SLP) as a programming environment for solving
decision problems from the class $\Sigma^P_2$. 

All these projects are supported by algorithms to process logic programs and 
by their implementations. Niemel\"a and Simons \cite{ns95,ns96} developed a
system, {\em smodels}, to compute stable models of logic programs. 
Algorithms to process DATALOG programs under parallel circumscription
were presented in \cite{cp98b}. A system DeReS to process a wider class of
programs -- default theories -- is described in \cite{cmt96,cmmt98}.
The emergence of implementations is the main reason why these
alternative logic programming systems are becoming viable computational tools.  

This paper is organized as follows.  In the next section we provide a
perspective of Horn logic programming, pointing to those of its features
that are responsible for its expressivity. We discuss the effects of
the negation operator in logic programming and show how it increases the
expressivity beyond the currently accepted bounds. In Section
\ref{stable-new} we formally introduce stable logic programming (SLP). 
Next, we study the expressive power of SLP, showing that the
applicability of SLP is attuned to the class NP. In Section \ref{rvc} we
note the limitations in using recursion in SLP and show that SLP
programs should be interpreted as descriptions of sets of constraints. 
Consequently, the methodology of programming with SLP is different from
that of ordinary logic programming. We discuss several examples and
show how SLP programs that solve them can be developed. We discuss the issue 
of uniform control associated with SLP in Section \ref{uniform}.
Conclusions and a ``road map'' for the future complete the paper.

\section{Horn logic programming}
\label{hlp}

The idea to use logic as a computational mechanism can be traced
back to the Herbrand's analysis of the effectivity of proofs in 
first-order logic \cite{her31}. In this work Herbrand discovered a
{\em unification algorithm}, one of the basic constructions behind 
the present-day automated deduction and logic programming. 

Transforming logic into a viable programming tool required two
additional crucial steps. First of them was the introduction of 
the {\em resolution} by Robinson \cite{ro65} with unification as one of
its key components. Over the years resolution was extensively studied 
\cite{cl73,lov78,mwa88} and gained in stature as one of the most 
successful techniques of automated reasoning.

The second key step was to narrow down the focus of automated deduction
to the class of {\em Horn theories} and even more specifically, to the
class of {\em definite} Horn theories. A {\em definite Horn clause} is a
formula of the form
\[
q_1(\vec{X})\wedge \ldots\wedge q_k(\vec{X})\Rar p(\vec{X})
\]
(written $p(\vec{X}) \lar q_1(\vec{X}),\ldots,q_k(\vec{X})$ by 
the logic programming
community), where $q_i$ and $p$ are atoms of the language. A definite 
Horn theory (a {\em Horn logic program}) is a finite theory consisting 
of definite Horn clauses. 

Each definite Horn theory $P$ is consistent and possesses a least 
Herbrand model, $LM(P)$. This least model provides a natural and
intuitive semantics for 
definite Horn theories and leads to a natural concept of computability. 
We say that a finite definite Horn theory $P$ 
{\em specifies} a subset $X$ of the {\em Herbrand universe} of $P$, 
$HU(P)$, if for some predicate symbol $p$ occurring in $P$ we have:
\begin{equation}\label{eq-1}
X = \{t\in HU(P)\colon p(t)\in LM(P)\}.
\end{equation} 
One of the fundamental results underlying the area of logic programming
is that every recursively enumerable set can be specified by a definite
Horn theory \cite{sm68,an78}. This result shows that definite Horn 
theories are as expressive as Turing machines and precisely capture 
the concept of Church-Turing computability.

The area of logic programming was born in early 70s when it was realized 
that the elements of the set $\{t\in HU(P)\colon p(t)\in LM(P)\}$ can
actually be {\em computed}. Namely, a special form of resolution, the 
{\em SLD-resolution} \cite{ko74,ave82}, and the so-called {\em lifting 
lemma} \cite{llo84} allow us to compute ground terms $t$ (more generally, 
ground  substitutions $\sigma$) such that $p(t)$ (or $p(\vec{X}\sigma)$) 
belongs to the least Herbrand model of a definite Horn theory. 

Availability of this uniform control mechanism, SLD-resolution, is the
key aspect of Horn logic programming. Horn programs, that is, definite 
Horn theories, need only to specify definitions and properties of objects 
and domains of interest. The programmer no longer needs to specify the 
exact way in which to perform the computation. The  control is provided 
by the mechanism of the SLD resolution. This feature of logic programming, 
the separation of logic (``what'' part) from control (``how'' part), was
and still remains one of the most attractive features of logic programming. 
It allows the programmers to focus only on the logic of the problem
and frees them from the burden of specifying the control. Consequently, 
it carries with itself a promise of easier code development, facilitates 
modular design of software and eases the problems of program 
verification\footnote{This expectation did not entirely materialize. 
Peculiarities of logic programming implementations, including subtleties 
of search space pruning and its side effects resulted in a paradigm not 
easily accepted by real-life programmers.}
\cite{ko79}. It is due to the separation of logic from control that 
logic programming is often classified as a {\em declarative} programming 
system.

Function symbols are critical for the Horn logic programming.
The presence of function symbols in the language, especially of
the list constructor $[\cdot]$, allows the programmer to encode 
hereditarily finite higher-order objects as terms of the language. 
Consequently, logic program clauses can be used to describe recursive 
definitions of higher-order objects. Modeling the recursive definitions 
within logic programs is responsible for the expressive power of logic
programming and became one of its most important and most common
programming techniques. 

The situation changes drastically if function symbols (in particular,
$[\cdot]$) are not available. This formalism was extensively studied 
as a possible query language by the database community and is known there 
as DATALOG (see, for instance, \cite{ul88}). With no function symbols, 
the Herbrand universe of finite Horn programs is finite. Thus, the ability 
to represent higher-order objects in DATALOG is significantly
restricted.  Similarly, many recursive definitions can no longer be
modeled by DATALOG clauses (those which describe how more complex objects of
higher order can be constructed from simpler ones). Consequently, 
the expressive power of finite DATALOG programs is very limited. They can 
only express a proper subset of all polynomial-time computable queries 
\cite{acy91}. Thus, it is the presence of function symbols in the language 
that is responsible for the expressive power of Horn logic programming.

We will now summarize some of the key features of Horn logic programming.
\begin{enumerate}
\item The existence of a single intended model (a least Herbrand model)
yields the semantics for Horn logic programs and the notion of computability. 
Horn programs compute extensions of predicates in the least model.
\item Function symbols in the language allow the programmer to encode 
higher-order objects as terms and represent recursive definitions
of these objects by means of Horn clauses.
\item Horn programs can specify any recursively enumerable set. Thus,
Horn logic programming precisely captures the commonly accepted
notion of computability.
\item Horn logic programming is declarative due to the separation of
logic, represented by Horn programs, from uniform control, represented 
by the SLD resolution.
\end{enumerate}

\section{Negation in logic programming}

As observed above, from the point of view of the expressive power, the Horn 
logic programming is as powerful as any programmer might want it to be
--- it captures recursively enumerable sets. However, as a declarative system, 
it was not quite satisfying. The ability to describe intuitive declarative 
specifications of objects to be computed was significantly hampered by 
disallowing the negation operator from the bodies of clauses. This was 
recognized very early on in the development of the field. In particular, 
the negation operator has been available in PROLOG since its creation 
\cite{ckpr} and extensions of Horn logic programming with negation in 
the bodies of program clauses were studied since mid 1970s. 

In fact, the effort to extend Horn logic programming by allowing 
the negation operator in the bodies was among the strongest driving 
forces behind the development of the area in the past 25 years. The task 
is far from straightforward as adding negation implies that the existence 
of a unique least model (one of the fundamental features of the Horn 
logic programming paradigm) is no longer guaranteed. 

Proposals to address the problem can be divided into two classes. 
Proposals of the first type attempt to salvage the notion of a {\em
single} intended model at a cost of narrowing down the class of programs
or weakening the semantics. Apt, Blair and Walker \cite{abw87} introduced 
the notion of {\em stratification}, a syntactic restriction on logic 
programs with negation. They assigned to each stratified program a single 
intended model, a {\em perfect model}. In another approach, van Gelder, 
Ross and Schlipf \cite{vrs91} assigned to an {\em arbitrary} program a single 
intended 3-valued model, a {\em well-founded model}. These two proposals 
are still very much in the spirit of the Horn logic programming paradigm. 
Namely, in each of these approaches, a program specifies extensions of 
predicates in a single intended model --- perfect or well-founded, 
respectively. Moreover, suitably modified versions of resolution were 
proposed as a uniform processing mechanisms \cite{ssw95,ssw97}. 

However, both these approaches lead to problems with excessive expressive 
power. Apt and Blair \cite{ab89} proved that stratified
programs with finite number of strata specify precisely arithmetic sets.
This means that they are expressive beyond what is at present considered
computable. Since the well-founded model coincides with perfect model
for stratified programs, the same result applies to well-founded
semantics\footnote{In particular, the programmer that uses a logic
programming environment based on the well-founded semantics (for instance 
XSB) must understand that some programs specify complex sets for which 
computation will not terminate. The non-termination occurs in Horn logic 
programming as well, but a more complex semantics adds an additional layer 
of complexity to programmer's task.}. In fact, perfect models of locally 
stratified programs \cite{Prz86} specify an even bigger class of sets --- 
the class of hyperarithmetic sets \cite{bms91}. The same class of sets
is specified  by programs for which the well-founded model is total while, 
in general, well-founded semantics specifies the class of $\Pi_1^1$ sets 
\cite{sch91}. 

The common idea behind the proposals of the second type was to
distinguish among all models of a program a {\em collection} of
intended models rather than a single one. The class of {\em supported}
models was introduced by Clark \cite{cl78} and, almost  ten years later,
Gelfond and Lifschitz \cite{gl88} fine-tuned Clark's approach and
defined the class of {\em stable} models as candidates for intended models of
logic programs with negation. 

Our goal in this paper is to present a perspective of logic programming 
with stable model semantics and, thus, from now on we will focus on 
stable models only. Let us briefly recall the definition of a stable
model (for a detailed treatment, the reader is referred to \cite{mt93}). 
Let $P$ be a logic program and let $P_g$ be the grounding of $P$. A subset 
$M$ of the Herbrand base of $P$ is a {\em stable model} of $P$, if $M$ 
coincides with the least model of the {\em reduct} $P_g^M$ of $P_g$ with 
respect to $M$. The reduct $P_g^M$ is the program obtained from $P_g$ 
by removing from $P_g$ all clauses containing in the body a literal of 
the form $\n(p)$, for some $p\in M$, and by removing literals of the 
form $\n(p)$ from all of the remaining clauses. Directly from this
definition one can derive the following fundamental properties of stable
models. First, every stable model of a logic prgram $P$ is, indeed, a model 
of $P$. Furthermore, every stable model of $P$ is a minimal model of $P$
and a supported model of $P$. Finally, the family of stable models of a
program form an antichain.

The stable model semantics, from the day it was proposed, was the subject 
of some controversy. On one hand, it was commonly accepted that stable 
models provide the right semantics for logic programming with negation. 
There is an abundance of evidence to support this claim. For instance,
it is known that the stable model semantics coincides with the least 
model semantics on definite Horn programs. For stratified logic programs 
it coincides with the perfect model semantics. Further, when 
the well-founded model is total, it defines a unique stable model 
\cite{vrs91}. Finally, the well-founded model is the least three-valued 
stable model \cite{prz90}. In addition, as demonstrated in 
\cite{mt89c,bf91a,ka97}, logic programming with the stable 
model semantics is closely related to default logic of Reiter \cite{re80}, 
a commonly accepted formalism for knowledge representation. 

On the other hand, it was not clear how to reconcile the stable model 
semantics with the paradigm of Horn logic programming, as presented in
Section \ref{hlp}. Three problems are: lack of a single intended model 
(the notion of specification given by (\ref{eq-1}) requires modifications), 
increase in the expressive power beyond the accepted limits of computability, 
and inadequacy of resolution-based control.

The first problem can, to some degree, be overcome by means of the so
called {\em skeptical semantics}. Under this semantics, a ground atom is 
entailed by the program if it is true in {\em all} of its stable models. 
We say that a logic program with negation $P$ {\em specifies} a subset 
$X$ of $HU(P)$, if for some predicate symbol $p$ occurring in $P$ we have:
\[
X = \{t\in HU(P)\colon P\models_{\mathit skeptical} p(t)\}.
\]
However, at this point, problems with the expressive power reappear. 
Indeed, stratified programs have unique stable model, thus
skeptical semantics coincides with perfect model semantics for such
programs. Consequently the results on the expressivity of perfect
semantics for stratified programs apply in this situation.
Moreover, the problem to decide the existence of a stable model of 
a finite logic program is $\Sigma_1^1$-complete, while the problem to
decide the membership in all stable models is $\Pi_1^1$-complete
\cite{mnr92b}. Thus, both problems are far beyond accepted notions of 
computability. 

There is one additional complication. Due to the complexity,
no form of resolution can be applicable in the general case of skeptical
semantics, without drastic restrictions on the syntax of programs. Whereas 
in the case of well-founded semantics conditions limiting the complexity 
are known and resolution-based systems for well-founded semantics were 
developed \cite{ssw95,ssw97}, in the case of skeptical semantics such
results have yet to be established. 

\section{Stable logic programming}
\label{stable-new}

The difficulty of fitting logic programming with stable model semantics
into the paradigm of Horn logic programming, combined with an intuitive
appeal of stable models, makes us believe that the place of stable model
semantics in logic programming must be reexamined.
To this end, we propose below an alternative paradigm 
to that of the Horn logic programming, a paradigm consistent with the 
properties of stable models. We will refer to it as {\em stable
logic programming} (or, {\em SLP} for short).

While stable logic programming is in many aspects different from
Horn logic programming, at the most general level it shares with it
the key feature of separation of logic from control. Consequently, as 
other logic programming formalisms, stable logic programming is declarative.
The programmer specifies the problem at hand as a logic program. This 
program is then processed by a uniform control mechanism, thus solving 
the original problem. The differences are in the
syntax and semantics. These differences affect the control mechanism (it is
no longer the SLD resolution), as well as the expressive power of stable
logic programming and the corresponding programming methodology. 

We will now specify stable model programming and discuss these differences 
in more detail. First, we will restrict the syntax since, as we saw 
earlier, without any restrictions 
the expressive power gets out of hand. Trivially, the negation operator 
must remain in the language (there is no need for the stable model 
semantics without it). The other major source of complexity, function 
symbols, must be eliminated, however. Indeed, even under a restriction 
of stratification, in the presence of both the negation operator and 
function symbols in the language, the complexity grows beyond the limits 
of computability \cite{ab89}. Thus, in stable logic programming we 
adopt the language of logic programming that consists of denumerable 
collections of constant, variable and predicate symbols. In addition, we
will allow for the negation operator to appear in the bodies of program 
clauses. Finite programs in this language will be referred to as 
{\em SLP programs}. 

Due to the presence of negation, the existence of a least Herbrand model
is no longer guaranteed. The semantics of SLP programs will be defined
in terms of their stable models. Before we address this issue in more
detail, let us observe that the formalism of finite function symbol-free logic 
programs with negation, was extensively studied by the database community. 
The formalism is often referred to as $\Dn$ and several semantics for 
$\Dn$ were studied. Stratified version of $\Dn$ with perfect model 
semantics and with well-founded semantics received particular attention 
\cite{ahv95}. At the same time, $\Dn$ with stable 
model semantics has 
never drawn any significant interest in database community, mostly due 
to the fact that under 
the stable model semantics there is no guarantee of a single intended 
model that might be used to determine an answer to a query stated as a 
$\Dn$ program.
Skeptical stable semantics was not regarded as quite satisfactory
either as the set of atoms entailed under this semantics is not a model
of a $\Dn$ query.

The lack of a single intended stable model, perceived as a problem by
the logic programming and database communities, plays the key role
in stable logic programming. Under the stable 
model semantics, a finite SLP program can be viewed as a specification 
of a {\em finite family of finite sets}. Namely, we say that a finite SLP 
program $P$ {\em specifies} a family of sets $\cal X$ if for some 
$k$-ary predicate $p$,
\begin{equation}
\label{eq-3}
{\cal X} = \{\{(c_1,\ldots,c_k)\in HU(P)\colon p(c_1,\ldots,c_k)\in
M\}
\colon \ \mbox{$M$ is a stable model of $P$}\}.
\end{equation}
This notion of specification is the counterpart, in the case of SLP, of
the notion of specification given by (\ref{eq-1}) in the case of Horn 
logic programming. It allows the programmer, without resorting to function 
symbols, to write logic programs that specify second-order objects which, 
in Horn logic programming, would be encoded by means of terms involving
the $[\cdot ]$ operator. Thus, as we will state it more formally later, 
multiple intended models allow us to recover in SLP some of the expressive 
power of logic programming lost by eliminating function symbols from 
the language. 

In addition, the notion of specification given by (\ref{eq-3}) suggests
that SLP programs are very well suited to represent problems whose 
solutions are finite families of finite sets. For instance, a
hamiltonian cycle in a directed graph is a set of edges, that is, a set 
of pairs of vertices. The collection of all hamiltonian cycles of a graph 
is then a collection of sets of pairs, that is an object of the form given by 
(\ref{eq-3}). In Section \ref{rvc}, we will exhibit explicit SLP programs
that represent the hamiltonian cycle problem. 

\section{Expressive power of SLP}\label{express}

Following Garey and Johnson \cite{gj79}, we define a search problem,
$\Pi$, to consist of a set of finite {\em instances}, $D_\Pi$. Further, 
for each instance $I \in D_\Pi$, there is a finite set $S_\Pi(I)$ of all
{\em solutions} of $\Pi$ for the instance $I$. An algorithm solves a
search problem if for each instance $I\in D_\Pi$ it returns the answer 
``no'' when $S_\Pi(I)$ is empty, and
any solution $s\in S_\Pi(I)$, otherwise. Notice that all decision
problems can be viewed as special search problems: for every instance
$I$ of a decision problem, define $S_\Pi(I) = \{$``yes''$\}$, if $I$ is a
``yes'' instance of $\Pi$, and $S_\Pi(I) = \emptyset$, otherwise.
Notice also that with each search problem one can associate a decision
problem: given an instance $I\in D_\Pi$, decide whether $S_\Pi(I)=
\emptyset$. 

Several interesting search and decision problems can be associated with
stable logic programming. Consider a finite SLP program $P$. Clearly,
the Herbrand universe, the Herbrand base and the grounding of $P$ are all 
finite. Consequently, stable models of $P$ (if exist) are finite, too. 
It follows that the problem to compute, given a finite SLP program, 
its stable models is a search problem. Another search problem of interest 
is, given a finite SLP $P$ and an element $a$ of the Herbrand base $HB(P)$, 
to compute a stable model of $P$ containing $a$. An associated decision 
problem asks for the existence of stable models of a finite SLP program 
$P$. Other related decision problems ask whether a given element $a$ 
of the Herbrand base $HB(P)$ belongs to some (or all) stable models of $P$. 

To understand the expressive power of SLP, we need to study which search
and decision problems can be reduced to search and decision problems
associated with SLP. We will first consider the class of decision problems 
and restrict, for a moment, to propositional programs only. The following 
theorem \cite{mt88} plays a key role in our discussion.

\begin{theorem}
\label{jacm}
The problem to decide whether a finite propositional logic program has a
stable model is NP-complete.
\end{theorem}

Theorem \ref{jacm} implies that for every decision problem $\Pi$ in 
the class NP and for every instance $I\in D_\Pi$, there is a 
{\em propositional} program $P^I_\Pi$ such that
\begin{enumerate}
\item[(1)] $P^I_\Pi$ can be constructed in time polynomial 
in the size of $I$, and
\item[(2)] $\Pi$ has a solution for $I$ if and only if $P^I_\Pi$ has a stable
model.
\end{enumerate}

Thus, any decision problem $\Pi$ in NP can be solved by a uniform
control mechanism of deciding existence of stable models of logic
programs. Given an instance $I$ for which $\Pi$ must be decided, one may 
encode $\Pi$ and $I$ as a propositional logic program $P^I_\Pi$
(that can be constructed in time polynomial in the size of $I$) and decide 
whether $I$ is a 
``yes'' instance of $\Pi$ by deciding whether $P^I_\Pi$ has a stable 
model. Moreover, any decision problem that can be decided in this way is 
in NP. 

Consider now a search problem $\Pi$. Assume that for every instance
$I\in D_\Pi$, there is a propositional program $P^I_\Pi$ satisfying the
condition (1) and a stronger version of the condition (2):
\begin{enumerate}
\item[(2$'$)] there is a polynomially computable {\em one-to-one} function
$\sol_\Pi^I$ from
the class of stable models of $P^I_\Pi$ to $S_\Pi(I)$. 
\end{enumerate}

Search problems of this type can be solved in a similar way to the one
described earlier for the case of decision problems. Given an instance
$I\in D_\Pi$, one constructs the program $P^I_\Pi$, finds its stable
model $s$, computes $\sol_\Pi^I(s)$ and returns it as a solution to 
$\Pi$ for $I$. If no stable model exists, the answer ``no'' is returned.

To the best of our knowledge, it is an open problem to characterize
the class of search problems for which this approach can be used, that
is, the class of search problems $\Pi$ for which programs $P_\Pi^I$ 
satisfying conditions (1) and (2$'$) can be found. We saw that all
decision problems in the class NP can be solved in this way. Furthermore,
all search problems whose associated decision problems are in NP, that we 
considered so far, also can be dealt with in this way (some of 
the encodings will be discussed later in the paper). 

The approach presented above relies on encodings of decision and search
problems as problems involving existence or computation of stable models
of propositional programs. It is not entirely satisfactory as different
programs $P^I_\Pi$ are needed for each instance $I$ of a problem or, to
put it differently, the logic is not separated from data. We will now 
present yet another possible approach that takes advantage
of variables in the language. 

Consider a search problem $\Pi$. Assume that there exist:
\begin{enumerate}
\item[(1)] an effective encoding $edb_\Pi$ under which every instance
$I\in D_\Pi$ is represented as a database under some, fixed for all
instances from $D_\Pi$, relational database scheme,
\item[(2)] a finite SLP program, $P_\Pi$, such that for every instance $I$,
that there is a  polynomially computable one-to-one function
$\sol_\Pi^I$ from the class of stable models of $edb_\Pi(I) \cup P_\Pi$ to 
$S_\Pi(I)$.
\end{enumerate}
Then, $\Pi$ can be solved  for an instance $I$ by first constructing the 
program $edb_\Pi(I) \cup P_\Pi$, then by finding its stable model $s$
and, finally, by reconstructing from $s$ a solution $\sol_\Pi^I(s)$.

This approach is more elegant and more in the spirit of standard
programming. In this approach, $P_\Pi$ can be regarded as a program
(logic) for solving the problem $\Pi$, and the database $edb_\Pi(I)$ can 
be viewed as data. Thus, there is a clear separation of logic (uniform
over all possible instances to the problem) and data (encodings of
problem instances).

We say that a search problem can be {\em solved by a uniform SLP
program} if there exist an encoding $edb_\Pi$ and a program $P_\Pi$ 
satisfying (1) and (2). As in the case of two earlier approaches, 
the question is which search problems can be solved by a uniform SLP
program.  The following strengthening of Theorem \ref{jacm}
was first proved in \cite{sch91}.

\begin{theorem}
\label{schlipf}
A decision problem can be solved by a uniform SLP program if and only if
it is in the class NP.
\end{theorem}

Thus, for decision problems we have a complete answer. The problem
remains open for arbitrary search problems. Let us point out, though,
that all search problems, whose associated decision problems are in NP,
that we considered so far, can be so solved.

To summarize, due to the absence of function symbols, the expressive
power of stable logic programming is restricted as compared to Horn
logic programming. However, due to the
use of negation and the stable model semantics, some of the lost 
expressive power is recovered. 
SLP can capture all decision problems in NP and many (perhaps all) search 
problems whose decision versions are in NP. 

\section{Recursion versus constraints}\label{rvc}

The restrictions in the syntax of SLP, the change in the semantics 
and, consequently, the change in the notion of specification (from
(\ref{eq-1}) to (\ref{eq-3})) requires a different approach to
programming. Perhaps most importantly, the use of recursion is severely
restricted. A limited version of recursion is still available. Namely, 
recursive definitions of predicates (or rather of sets that are their 
extensions) can be modeled by SLP clauses. For instance, the following 
SLP clauses define the transitive closure of a relation $rel$ 
\begin{quote}
$tc(X,Y) \lar rel(X,Y)$\\
$tc(X,Y)\lar tc(X,Z),rel(Z,Y)$
\end{quote}
However, without function symbols to build terms representing higher-order
objects, it is far from clear how a clause could capture 
such recursive definitions that specify how more complex objects are 
constructed from simpler ones. For instance, consider the following
HLP program:
\begin{quote}
%(1)\hspace*{0.3in}$allconnected(Y,[X])\lar edge(Y,X)$\\
%(2)\hspace*{0.3in}$allconnected(Y,[H|T])\lar
%edge(Y,H),allconnected(Y,T)$\\
(1)\hspace*{0.3in}$clique([X])\lar vertex(X)$\\
(2)\hspace*{0.3in}$clique([Y|X])\lar clique(X),allconnected(Y,X)$
\end{quote}
We assume here that the predicate $allconnected(Y,X)$ is defined so that
to succeed precisely when vertex $Y$ is connected by an edge to all
vertices on list $X$, and when $X$ has no repetitions. The program
consisting of the definition of $allconnected$, of clauses (1) and (2)
and of the description of a graph $G$ in terms of facts specifying 
the extensions of the predicates $vertex$ and $edge$,
computes all cliques in $G$. Notice that each time
clause (2) is used in computation, it produces a longer list (more
precisely, each iteration of the one-step operator associated with 
this program generates new ground terms that need to be included
in the extension of the predicate $clique$. 
The capability of growing the set of available ground terms is both a 
strength and a weakness of HLP. It allows to code all hereditarily finite 
sets but, at the same time, makes it possible to write programs that do 
not terminate.

This phenomenon does not occur in SLP --- the available constants are 
prescribed from the beginning and no new terms can be built of these 
constants as there are no function symbols that could accomplish this.
At the same time, as we saw earlier, SLP is expressive enough to specify 
{\em some} higher-order objects. Namely, an SLP program specifies 
the collection of its stable models (or, as we defined in (\ref{eq-3}), 
collections of extensions of predicates in stable models). Clauses of 
the program do not represent recursive definitions of individual stable 
models but act as {\em constraints}. More specifically, a ground clause
\[
C = p\leftarrow q_1,\ldots,q_m,\n(r_1),\ldots ,\n(r_n)
\]
expresses the following constraint: at least one $q_i$ does not belong 
to a putative stable model $M$, or at least one $r_j$ belongs to $M$, or
$p$ belongs to $M$. In other words, $M$ must be a model of the clause
$C$ treated as a propositional formula. Moreover, in a crucial 
difference with propositional logic, stable logic programming adds
to this constraint also a preferred way to compute sets that satisfy it: 
in order to enforce the constraint, once all $q_i$ are computed and
none of $r_j$ was established, $p$ is added to the set rather than some
$q_i$ eliminated or some $r_j$ added. 

Here lies the key difference between Horn logic programming and stable
logic programming. To specify a higher-order object in Horn logic 
programming, the programmer models the object as a term and represents in 
a program a recursive definition of the object. To specify a second-order 
object in SLP, the programmer thinks of the object as a stable model of 
a program and constructs the program by modeling a definition of 
the object expressed in terms of constraints. This feature makes 
SLP especially well suited to deal with constraint satisfaction
problems, a point made in \cite{nie98}. 

An important issue raised by our discussion is how to represent
constraints by SLP clauses. We will now present one such technique 
of adding {\em selection} clauses (in the case of default logic, this
techniques is discussed in \cite{cmmt98}). 

Many applications specify objects of interest to a problem at hand by first
specifying a general domain from which these objects have to be selected (the
family of all subsets of a set, the family of all $n$-tuples over a given
set, etc.) and, then, by specifying additional conditions (constraints)
these objects have to satisfy. In particular, constraint satisfaction 
problems are of this type.

Thus, to develop programs encoding solutions to such problems we might 
proceed in two steps. First, we develop an SLP program whose family of 
stable models encodes the general domain of candidate objects (all 
subsets, all sequences of length $n$, etc.). Next, we add to this
program clauses representing constraints that must be enforced.

In many applications involving subsets of some universe $U$ constraints are 
of the following form (here, $A$ and $B$ are two subsets of the same
universe $U$):
\begin{description}
\item[$C(A,B)$] if a solution contains the set $A$, it must also contain
at least one element from $B$ (or, in other words, a solution must not 
contain the set $A$ or must contain at least one element from the set
$B$)
\end{description}
Assume that $P$ is an SLP program whose stable models are subsets of
$U$. Let $\cal S$ be a family of all stable models of $P$. To enforce
the constraint $C(A,B)$ on sets in $\cal S$, that is, to construct 
a program whose stable models are exactly these stable models of $P$ 
that satisfy $C(A,B)$, it is enough to add to $P$ the clause
(here $A = \{a_1,\ldots,a_k\}$ and $B = \{b_1,\ldots,b_m\}$):
\begin{quote}
$k_{C(A,B)}$:\hspace*{0.3in}$f\lar 
a_1,\ldots,a_k,\n(b_1),\ldots,\n(b_m),\n(f)$
\end{quote}
where $f$ is an atom not occurring in $P$. Formally, we have the
following theorem.

\begin{theorem}
\label{kill}
Let $P$ be a logic program and let $\cal S$ be the family of stable
models of $P$. Then, for every constraint $C(A,B)$ where $A$ and $B$ are 
sets of atoms, $M$ is a stable model of $P\cup \{k_{C(A,B)}\}$ if and 
only if $M\in {\cal S}$ and $M$ satisfies $C(A,B)$.
\end{theorem}

As a corollary from Theorem \ref{kill}, it follows that conjunctions of 
constraints of type $C(A,B)$ (that is, formulas in the conjunctive normal
form) can be enforced by SLP programs consisting of clauses
$k_{C(A,B)}$, for each conjunct $C(A,B)$. Moreover, the size of this SLP
program is linear in the size of the CNF formula.

Theorem \ref{kill} is the first step towards a methodology of
programming with SLP programs. A {\em systematic} study of this
and other programming techniques appropriate in the case of SLP has not 
yet been conducted and has to be performed before SLP is used as 
a programming language. However, even this technique alone is quite
powerful. We will now illustrate how it can be applied to encode several
combinatorial problems.

First, we will revisit our ``clique'' problem. We assume that the graph
is described by two lists of facts: $vertex(a)$ for all vertices $a$ of the
graph (we will denote the set of vertices by $V$), and $edge(a,b)$, for 
all edges $\{a,b\}$ of the graph. We denote these lists of facts by
$edb_{clq}(G)$. To specify 
cliques by an SLP program, one first needs to write a program specifying 
all subsets of the set of vertices of a graph. This can be accomplished, 
for instance, by the following two clauses:

\begin{quote}
(CLQ1)\hspace*{0.3in}$in (X) \lar vertex(X),\n(out(X))$\\
(CLQ2)\hspace*{0.3in}$out(X) \lar vertex(X),\n(in(X))$
\end{quote}
Stable models of the program consisting of clauses (CLQ1) and (CLQ2), and
of facts $edb_{clq}(G)$ are of the form
\[
edb_{clq}(G)\cup \{in(v)\colon v\in K\}\cup\{out(v)\colon v\notin K\},
\]
where $K$ is a subset of the vertex set. Thus, the stable models of 
this program are in one-to-one correspondence to all subsets of
the vertex set and can be regarded as their representations.

Next, we need to select those stable models
that represent sets satisfying the clique condition: any two vertices in 
a clique are connected by an edge. This condition may be expressed as a
constraint of the type $C(A,B)$: if two vertices are in a clique then
they are equal or they are connected with an edge. Thus, all these 
constraints (for all pairs of vertices) can be expressed by a single 
SLP clause:
\begin{quote}
(CLQ3)\hspace*{0.2in}$f\lar 
vertex(X),vertex(Y),in(X),in(Y),\n(x = y),\n(edge(x,y)),\n(f)$
\end{quote}
By Theorem \ref{kill}, the program consisting of facts $edb_{clq}(G)$, and
clauses (CLQ1) - (CLQ3) has 
as its stable models precisely those sets of the form $edb_{clq}(G)\cup 
\{in(v)\colon v\in
K\}\cup\{out(v)\colon v\notin K\}$, for which $K$ is a clique. 

We will now further illustrate this approach by describing programs
encoding the following problems:
\begin{enumerate}
\item Computing hamiltonian cycles in directed graphs,
\item Computing models of a propositional CNF formula,
\end{enumerate}

For the first problem, we will need to represent directed graphs. Let
$G$ be a directed graph with the set of vertices $V$ and with the set of
directed edges $E$. We will represent $G$ by the following facts: 
$vertex(a)$, for all vertices $a\in V$, $edge(a,b)$, for all 
{\em directed} edges $(a,b)\in E$, and $initialvtx(a0)$, for some
vertex $a0\in V$. We will denote this representation of $G$ by
$edb_{ham}(G)$.

Next consider the following clauses:
\begin{quote}
(HAM1)\hspace*{0.3in}$in(V1,V2)\lar edge(V1,V2), \n(out(V1,V2))$\\
(HAM2)\hspace*{0.3in}$out(V1,V2)\lar edge(V1,V2), \n(in(V1,V2))$
\end{quote}
These two clauses, together with the set of facts $edb_{ham}(G)$ define 
an SLP program
whose stable models are of the form 
\[
edb_{ham}(G)\cup\{in(a,b): (a,b)\in A\}\cup \{out(a,b): (a,b) \notin A\},
\]
for some set of edges $A$. Thus, these
stable models represent all subsets of the set of edges of $G$. Those
sets of edges that are hamiltonian cycles satisfy the following
additional constraints: (1) if two edges of the cycle end in the same vertex, then 
these edges are equal, and (2) if two edges of the cycle start in the
same vertex then these edges are equal. By Theorem \ref{kill}, these 
constraints can be enforced by adding the following two clauses:
\begin{quote}
(HAM3)\hspace*{0.3in}$f\lar in(V2,V1),in(V3,V1),\n(V2=V3),\n(f)$\\
(HAM4)\hspace*{0.3in}$f\lar in(V1,V2),in(V1,V3),\n(V2=V3),\n(f)$
\end{quote}
The stable models of the so expanded program are of the form
$edb_{ham}(G)\cup\{in(a,b): (a,b)\in A\}\cup \{out(a,b): (a,b) \notin A\}$, 
for some set of edges $A$ spanning in $G$ a set of vertex disjoint paths
and cycles. Adding the following two clauses
\begin{quote}
(HAM5)\hspace*{0.3in}$reached(V2) \leftarrow in(V1,V2),reached(V1)$\\
(HAM6)\hspace*{0.3in}$reached(V2) \leftarrow in(V1,V2),initialvtx(V1)$
\end{quote}
expands each of these stable models by the set of atoms $reached(a)$,
for all these vertices $a$ that can be reached from the vertex $a0$ by
means of a nonempty sequence of edges that are ``in'' the model 
(note that we use this weaker notion of recursion here, that is still 
available in SLP). A stable model encodes a hamiltonian cycle if all
vertices are reached. This constraint is enforced by the clause:
\begin{quote}
(HAM7)\hspace*{0.3in}$f \leftarrow \n(reached(X)), \n (f )$
\end{quote}
It follows that the program consisting of clauses (HAM1) - (HAM7) and of the
facts in the set $edb_{ham}(G)$ has as its stable models the sets of the form
\[
edb_{ham}(G)\cup\{reached(a): a\in V\}\cup\{in(a,b): (a,b)\in A\}\cup 
\{out(a,b): (a,b) \notin A\},
\]
for which $A$ is the set of the edges of
a hamiltonian cycle in $G$.

For the satisfiability problem, we need to represent CNF formulas.
Consider a CNF formula $\vph$ with the set of clauses $\cal C$ and the
set of variables $V$. The formula $\vph$ will be represented by several 
lists of facts: $var(a)$, for each $a\in V$, $clause(c)$, for each
clause $c\in{\cal C}$, $pos(c,v)$, for each variable $v$ and clause
$c$ such that $v$ appears positively in $c$, and $neg(c,v)$, for each
variable $v$ and clause $c$ such that $v$ appears negatively in $c$.
This specification of a CNF formula $\vph$ will be denoted by
$edb_{sat}(\vph)$. 

Consider now the following clauses:
\begin{quote}
(SAT1)\hspace*{0.3in}$true(X) \leftarrow\ \  var(X),\n (false (X))$\\
(SAT2)\hspace*{0.3in}$false(X) \leftarrow\ \ var(X),\n (true (X))$
\end{quote}
These two clauses generate all possible truth assignments to variables
in $V$. More formally, the program consisting of $edb_{sat}(\vph)$ and
clauses (SAT1) and (SAT2) has as its stable models the sets of the form
$edb_{sat}(\vph)\cup\{true(v):v\in U\}\cup \{false(v): v\in V\setminus
U\}$, where $U$ is a subset of $V$. Thus, they represent the set
of all valuations. The next two clauses
\begin{quote}
(SAT3)\hspace*{0.3in}$sat (C) \lar var(X), clause(C), true(X), pos(C,X)$\\
(SAT4)\hspace*{0.3in}$sat (C) \lar var(X), clause(C), false(X), neg(C,X)$
\end{quote}
simply define when a clause is satisfied and add to each stable model
the set of clauses that are true in the valuation represented by this
model. Formula $\vph$ is satisfiable if there is a valuation which makes
all clauses true. This requirement can be enforced by adding the clause
\begin{quote}
(SAT5)\hspace*{0.3in}$f \leftarrow\ \  clause(C), \n(sat(C)), \n (f)$
\end{quote}
It follows that $M$ is a stable model of the program consisting of
clauses (SAT1) - (SAT5) and of the facts in the set $edb_{sat}(\vph)$ if and
only if \[
M = edb_{sat}(\vph)\cup \{sat(c): c\in{\cal C}\}\cup
\{true(v):v\in U\}\cup \{false(v): v\in V\setminus U\},
\]
where 
$U\subseteq V$ is a (propositional) model of $\vph$.

The encodings presented so far are {\em uniform}. That is, input data to
a problem is encoded as a collection of facts and the constraints defining 
the problem as clauses (usually with variables) with the latter part
not depending on a particular input. These two parts correspond well to 
the extensional and intensional components of a $\Dn$ program. However,
other encodings are also possible (and often easier to come up with). 

To make the point, let us again consider the satisfiability problem.
As before, consider a CNF formula $\vph$ with a set of variables $V$ and
a set of clauses $\cal C$. This time we will represent valuations as
subsets of $V$. It turns out that there is a very simple
propositional encoding of the satisfiability problem. First, note that
the clauses 
\begin{quote}
(SAT$'$1)\hspace*{0.3in}$in(v)\lar \n(out(v))$\\ 
(SAT$'$2)\hspace*{0.3in}$out(v)\lar \n(in(v))$
\end{quote}
(where in both (SAT$'$1) and (SAT$'$2) $v$ ranges over $V$),
specify all subsets of $V$. That is, the stable models of the program 
consisting of all clauses (SAT$'$1) and (SAT$'$2) are precisely the sets
of the form $\{in(a): a\in M\}\cup \{out(a): a\in V\setminus M\}$, where 
$M$ is a subset of $V$. Now, for each clause
\begin{quote}
$c$:\hspace*{0.3in}$\neg a_1\vee\ldots\vee\neg a_k\vee b_1\vee\ldots \vee
b_m$
\end{quote}
of $\vph$, add to the program the SLP clause
\begin{quote}
(SAT$'$3)\hspace*{0.3in}$f\lar a_1,\ldots,a_k,\n(b_1),\ldots,\n(b_m),\n(f)$
\end{quote}
Since a set of atoms satisfies clause $c$ if and only if it satisfies
the constraint $C(\{a_1,\ldots,a_k\}$, $\{b_1,\ldots,b_m\})$, it follows
that the program consisting of all clauses (SAT$'$1) - (SAT$'$3) has as its stable
models sets of the form $\{in(a): a\in M\}\cup \{out(a): a\in V\setminus
M\}$, where $M$ is a model of $\vph$.

Let us emphasize that the encodings discussed in this section are 
not unique. Two satisfiability encodings given here constitute but one 
example. Further, a different encoding of the hamiltonian cycle problem 
can be found 
in \cite{nie98} and more encodings (as propositional default theories and
logic programs) for several combinatorial problems were given in 
\cite{cmmt98}. These encodings may have different computational 
properties. In particular, the second encoding of satisfiability, even
though it is not uniform and requires that a separate program be created
for each satisfiability instance, may actually be better suited for
processing. Thus, the issue of the programming methodology likely to result in
programs whose stable models can be quickly computed is very important.
It has not been studied yet but must receive significant attention if
stable logic programming is to become a practical problem solving tool.

\section{Uniform control in SLP}\label{uniform}

As we have seen in the earlier sections, SLP programs can specify
a wide class of search and decision problems. In addition, they do so in
a declarative fashion by modeling, in a direct way, constraints defining 
a problem at hand. Thus, SLP programs are well suited to represent the
``logic'' part in Kowalski's ``algorithm = logic + control'' phrase.

In order for the stable logic programming to serve as an effective
computational problem solving tool (and not only as a knowledge
representation formalism), we need to develop the other component of 
the Kowalski's equation --- a {\em uniform control}. Since, in stable
logic programming, problems are encoded by SLP programs and solutions
correspond to stable models, this uniform control must consist of
algorithms to process SLP programs and compute (or decide the existence of) 
their stable models.

Several such algorithms were proposed in the recent years. In particular, 
algorithms were proposed to decide the existence of stable models of 
a {\em propositional} logic program $P$, and to compute one (or all) stable 
models of $P$, if they exist. Algorithms that decide the membership of 
an atom in some or all stable models were also developed 
\cite{ns96,cmt96,elmps97,adn97}. 

These algorithms employ a backtracking search through the space of all 
subsets of the Herbrand base of the program (the collection of all 
propositional letters that occur in the program). They also use a variety 
of search space pruning techniques. Some of these techniques rely on 
a generalization of the concept of stratification \cite{cmt96}. Other 
methods use well-founded semantics in a way unit propagation is used
in Davis-Putnam procedure for computing models of CNF formulas
\cite{ns96}.

Notice that since there are no function symbols in the language of SLP, 
the Herbrand universe, the Herbrand base and the grounding of an
{\em arbitrary} (that is, not necessarily propositional) finite SLP program 
are finite. There are straightforward algorithms to produce the (finite) 
ground version of a finite SLP program. More sophisticated algorithms,
minimizing the size of the ground program while preserving the stable
models, were proposed recently in \cite{ns96,cho96}. 
Thus, algorithms to compute 
stable models of finite propositional programs can be used with arbitrary 
finite SLP programs, too (and termination is guaranteed).

These algorithms led to implementations of several systems for processing 
SLP programs and for computing their stable models. Among them are
{\em smodels} \cite{ns95,ns96}, DeReS \cite{cmt96} and the system described 
in \cite{elmps97}. Any of these systems can be used as a control mechanism for 
the stable logic programming environment and transforms SLP from a
knowledge representation formalism into a computational programming tool. 

Thus, in the transition from Horn logic programming to SLP not only the
semantics and the methodology of programming changes. The control has
to change, too. Instead of SLD resolution, the basis for the control
mechanism of SLP is provided by backtracking search algorithms for
search spaces of subsets of a given finite set (Herbrand base of a
finite SLP program).

The stable logic programming became a viable proposal for a new logic
programming system with recent advances in algorithms for computing stable 
models and subsequent implementations of these algorithms 
\cite{ns96,cmt96,elmps97}. Even though comprehensive studies of these
implementations have yet to be performed, available results
provide reasons for optimism. For some classes of programs, systems 
such as {\em smodels} \cite{ns96} and DeReS \cite{cmt96} can successfully 
process programs with tens of thousands of clauses. In addition, as reported 
in \cite{nie98}, {\em smodels} can successfully compete on a class of
planning problems with special purpose planners (for more comprehensive
discussion of applications of logic programming to planning see
\cite{li98} and a collection of papers referred to there). It is clear that
with the performance of the systems computing stable models
improving, the attractiveness of SLP as a computational tool will grow.

Despite a significant progress and the existence of systems such as {\em
smodels} and DeReS, much more work on algorithms for computing stable
models is needed in order to obtain acceptable performance. There are 
several open problems such as development of new and more powerful pruning
techniques and study of probabilistic algorithms for stable model 
computation. 

\section{Conclusions and future directions}

In the paper we discussed the stable model semantics as the foundation 
of a computational logic programming system different from Horn logic
programming. This system, the stable logic programming or SLP, shares with 
other logic programming systems their key feature: the separation of 
logic from control. However, despite the fact that the stable model
semantics has its roots in the efforts to extend the principles of Horn
logic programming to the case of programs with negation, the stable
logic programming in several aspects differs significantly from standard 
logic programming systems. 
\begin{enumerate}
\item In the SLP programs are assigned a collection of intended models
rather than a single intended model as in Horn logic programming,
stratified logic programming or logic programming with well-founded
semantics.
\item Since there are no function symbols in the language, higher-order
objects are represented in the SLP as stable models of programs rather
than as ground terms of the Herbrand universe, as it is the case in Horn
logic programming and other similar systems.
\item SLP programs are interpreted as sets of constraints on objects to
be computed unlike in Horn logic programming where clauses model
recursive definitions rather than constraints. 
\item The control mechanism of SLP is no longer resolution-based.
Instead, the uniform control of SLP consists of backtracking search
algorithms for computing stable models of programs.
\item The SLP has lower expressive power than standard logic programming
systems. While it may seem to be a limitation, the class of problems
that can be solved in the SLP is still quite wide and includes all
decision problems from NP and many search and constraint satisfaction
problems of importance in artificial intelligence and operations
research. 
\end{enumerate}

We believe that the perspective of stable logic programming presented
in the paper certainly warrants further investigations. We will now
outline several interesting research directions. 

Let us start with the following fundamental question: why to use
the stable logic programming and not simply propositional logic.
Indeed, a finite collection of clauses (possibly with variables but not 
with function symbols) together with a finite collection of facts
can be viewed as a representation of a finite propositional formula in
the conjunctive normal form. Consequently, it can be viewed as an
encoding of the family of models of this formula (in the same way in
which an SLP program represents the family of its stable models - a key
observation underlying stable model programming). Moreover, since
propositional satisfiability problem is NP-complete, all decision
problems in NP can be reduced in polyniomial time to satisfiability
testing (and solved by means of programs such as Davis-Putnam procedure
for satisfiability testing). Finally, constraints of the type $C(A,B)$ 
can easily be encoded as clauses.

Is there then a real need to resort to logic programming? This is a
challenging open problem. We believe the answer is positive but cannot
offer any rigorous argument in support of the claim. In our
opinion, the advantage of stable logic programming stems from the
following two properties of stable models. First, they are minimal. 
Thus, when dealing with optimization problems minimality comes for free 
with the stable model semantics. Second, they are {\em grounded} --- 
facts are included in stable models only if they can be justified. 
Furthermore, a comparison between existing encodings of combinatorial 
problem shows that in many cases encodings in terms of logic programs are 
more concise than those in terms of satisfiability (possibly due to 
the groundedness property of stable models). In particular, we believe
that the most concise encodings of the existence of a hamiltonian cycle
problem as the problem of the existence of a stable model of a logic program
are asymptotically more more concise than similar encodings as the
problem of the existence of a satisfying valuation of a CNF formula.

Developing a formal setting to compare stable logic programming with 
propositional logic and providing a rigorous account of advantages and
disadvantages of both approaches are important general theoretical 
challenges. 

The second important research direction is the development of a systematic 
study of the methodology for developing SLP programs. Some initial steps in 
this direction were presented in Section \ref{rvc}.
There is a potential trade-off here. On one hand, one of the most
important objectives is ease of program development. On the other hand, 
we want our programs to run fast.

Third, despite the recent successes with implementing systems based on
the stable model semantics, there is still much to be done. So far, 
performance studies for the existing systems have been rather ad hoc. 
More comprehensive experimental studies are needed that will give
insights into the computational nature of stable models and will lead 
to faster algorithms. To support such studies one needs benchmarking systems.
One step in this direction is TheoryBase, a system described in 
\cite{cmmt94,cmmt98}.

Next, the SLP seems to be especially well suited for dealing with
constraint satisfaction problems. Thus, it is important to extend the
language of the SLP so that important classes of constraints involving
arithmetic operations and relations became easier to model. Possibility
of incorporating the SLP into existing constraint solving systems is
another important problem.

Finally, there is already evidence that the SLP can be a useful tool in
solving planning problems \cite{nie98}. Studying applicability of the
SLP paradigm to other classical problems of artificial intelligence and
operations research  may provide additional motivation to focusing on
this approach and may gain badly needed recognition to logic programming.

%\bibliographystyle{alpha}
%\bibliography{/a/al/u/d5/csfac/mirek/logic/nonmonlog}

\newcommand{\etalchar}[1]{$^{#1}$}

\end{document}